\documentclass[12pt]{iopart}
\usepackage{amsfonts}
\usepackage{iopams}  
\usepackage{setstack}  
\usepackage{cite}
\usepackage{amsthm}
\usepackage{graphicx}

\newcommand{\parti}[2]{\frac{\partial #1}{\partial #2}}

\newcommand{\diff}[2]{\frac{d #1}{d #2}}

\newcommand{\intall}{\int_{-\infty}^{\infty}}

\newcommand{\abs}[1]{\left|#1\right|}
\newcommand{\bk}[1]{\left(#1\right)}
\newcommand{\Bk}[1]{\left[#1\right]}
\newcommand{\BK}[1]{\left\{#1\right\}}

\newcommand{\trace}{\textrm{tr}}

\begin{document}
\title{Optomechanical parameter estimation}

\author{Shan Zheng Ang$^1$, Glen I.\ Harris$^2$, Warwick P.\ Bowen$^2$, and Mankei Tsang$^{1,3}$}

\address{$^1$ Department of Electrical and Computer Engineering, National University of Singapore, 4 Engineering Drive 3, Singapore 117583}

\address{$^2$ Centre for Engineered Quantum Systems, University of Queensland, St. Lucia, Queensland 4072, Australia}

\address{$^3$ Department of Physics, National University of Singapore, 2 Science Drive 3, Singapore 117551}

\ead{eletmk@nus.edu.sg}

\date{\today}

\begin{abstract}
  
  We propose a statistical framework for the problem of parameter
  estimation from a noisy optomechanical system. The Cram\'er-Rao
  lower bound on the estimation errors in the long-time limit is
  derived and compared with the errors of radiometer and
  expectation-maximization (EM) algorithms in the estimation of the
  force noise power. When applied to experimental data, the EM
  estimator is found to have the lowest error and follow the
  Cram\'er-Rao bound most closely. Our analytic results are envisioned
  to be valuable to optomechanical experiment design, while the EM
  algorithm, with its ability to estimate most of the system
  parameters, is envisioned to be useful for optomechanical sensing,
  atomic magnetometry, and fundamental tests of quantum mechanics.
\end{abstract}

\maketitle
\section{Introduction}

There has been spectacular technological advances in the use of
high-quality optomechanical oscillators for force sensing, enabling
ultra-sensitive force measurements of single spin, charge,
acceleration, magnetic field and mass~\cite{Rugar04_Nat,
  Cleland98_Nat, Krause12_NatPhot, Forstner12_PRL,
  Jensen08_NatNano}. Such advances have opened up the exciting
possibility of experimentally studying quantum light-matter
interactions in macroscopic structures~\cite{Purdy13_Sci,
  Verhagen12_Nat}, hence paving the way towards new technologies for
quantum information science and metrology~\cite{Mancini03_PRL,
  Brooks12_Nat,purdy2013,safavi2013}.

Thermal and measurement noises impose major limitations to the
accuracy of mechanical force sensors. Furthermore, while the
development of higher quality and lower mass mechanical oscillators
has played a central role in advancing the sensitivity of
optomechanical force sensors \cite{gavartin}; such oscillators also
have increased sensitivity to their environment.  This introduces new
sources of noise that can cause fluctuations in parameters, such as
the effective oscillator temperature and mechanical resonance
frequency. As optomechanical technology continue to advance, it can be
expected that methods to characterize, monitor, and control these
additional noise sources, in conjunction with thermal and measurement
noise, will become increasingly important. In this context,
statistical signal processing techniques that are provably optimal in
a theoretical sense offer the potential to improve the actual sensing
performance significantly, beyond the heuristic curve-fitting
procedures commonly employed in the field.

In this paper, we introduce a statistical framework to study the
problem of parameter estimation from a noisy optomechanical
system. This problem is especially relevant to the recent
optomechanics experiments reported in
Refs.~\cite{gavartin,harris_bowen2013}. We derive analytic
expressions for the Cram\'er-Rao lower bound on the estimation errors
and apply various estimation techniques to experimental data to
estimate the parameters of an optomechanical system, including the
force power, mechanical resonance frequency, damping rate, and
measurement noise power.

Our analytic results provide convenient expressions of the estimation
errors as a function of system parameters and measurement time and
should be valuable to optomechanical experiment design.  Another
highlight of our study is the use of the expectation-maximization (EM)
algorithm \cite{dempster, shumway_stoffer, levy}, which is generalized
here for a complex Gauss-Markov model and applied to a cavity
optomechanical system, both for the first time to our knowledge.
Among the estimators we have studied, including the one used in
Refs.~\cite{gavartin,harris_bowen2013}, we find that the
root-mean-square errors of the EM algorithm in estimating the force
noise power are the lowest, following the Cram\'er-Rao bound most
closely and beating the estimator in
Refs.~\cite{gavartin,harris_bowen2013} by more than a factor of 5 for
longer measurement times.

Our framework is also naturally applicable to quantum systems that can
be described by a homogeneous Gauss-Markov model
\cite{wiseman_milburn}, such as quantum optomechanical systems
\cite{chen2013, aspelmeyer2013,wheatley, yonezawa,iwasawa} and atomic
spin ensembles \cite{stockton,petersen}.  This makes our study, and
the EM algorithm in particular, relevant not just to future precision
sensing and system identification applications, but also to
fundamental tests of quantum mechanics
\cite{chen2013,aspelmeyer2013,testing_quantum}.

\section{Experiment}
 
To motivate our theoretical model and numerical analysis, we first
describe the optomechanical experiment presented in
Ref.~\cite{harris_bowen2013} that was used to produce the data.
The transducer under consideration consists of a room temperature
microtoroidal resonator that simultaneously supports mechanical modes
sensitive to external forces and high quality optical modes that
permit ultra-precise readout of the mechanical displacement.
We couple shot-noise limited 1550nm laser light into a whispering
gallery mode of the microtoroid via a tapered optical fiber which is
nested inside an all fiber inteferometer. Excitation of the mechanical
mode, which has fundamental frequency, damping rate and effective mass
of $\Omega_{\rm m}\!=\!40.33~\rm MHz$, $\gamma \!=\!23~\rm kHz$ and
$m_{\rm eff}\!=\! 7~\rm ng$ respectively, induces phase fluctuations
on the transmitted light which is measured by shot-noise limited
homodyne detection.
To maintain constant coupling of optical power into the microtoroid we
use an amplitude and phase modulation technique to actively lock the
toroid-taper separation~\cite{Chow12_OptExp} and laser frequency
respectively. The relative phase of the bright local oscillator to the
signal is controlled via a piezoactuated fiber stretcher that
precisely tunes the optical path length in one arm of the
inteferometer.
To specifically demonstrate power estimation, a small incoherent
signal is applied to the mechanical oscillator in addition to the
thermal fluctuations. This is achieved by the electrostatic gradient
force applied by a nearby electrode driven with white noise from a
signal generator~\cite{Lee10_PRL}.

The measurement record is acquired from the homodyne signal by
electronic lock-in detection which involves demodulation of the
photocurrent at the mechanical resonance frequency allowing real time
measurement of the slowly evolving quadratures of motion, denoted
$I(t)$ and $Q(t)$ where $x(t)\!=\!I(t)\cos(\Omega_{m}t) +
Q(t)\sin(\Omega_{m}t)$. The room temperature thermal fluctuations of
the mechanical mode are observed with a signal-to-noise ratio of
$37$dB and calibrated via the optical response to a known reference
modulation~\cite{Schliesser08}.  The resulting force sensitivity,
which can be extracted from Fourier analsis of the measurment record,
will depend on the specific protocol used.
Here we evaluate the force sensitivity of 3 parameter estimation
protocols relative to the Cram\'er-Rao lower bound.

\section{Theory}

\subsection{\label{model}Continuous-time model}
A simple linear Gaussian model for the mechanical mode can be
described by the following equation for the complex analytic signal
$z(t)$ of the mechanical-mode displacement:
\begin{eqnarray}
\diff{z(t)}{t} = -\gamma z(t)+i\Omega z(t) + \xi(t),
\label{a}
\end{eqnarray}
where $\Omega$ is the mechanical resonance frequency relative to
$\Omega_m$, $\gamma$ is the damping rate, and $\xi(t)$ is the
stochastic force as a sum of the thermal noise and the
signal. $\xi(t)$ is assumed to be a complex zero-mean white Gaussian
noise \cite{vantrees3} with power $A$ and covariance function
\begin{eqnarray}
E\Bk{\xi(t)\xi^*(t')} = A\delta(t-t'),
\qquad
E\Bk{\xi(t)\xi(t')} = 0.
\label{xi}
\end{eqnarray}
The measurements can
be modeled in continuous time as
\begin{eqnarray}
y(t) = C z(t) + \eta(t),
\label{y}
\end{eqnarray}
where $C$ is a real parameter and $\eta(t)$ is the measurement noise,
assumed to be a complex additive white Gaussian noise with power $R$:
\begin{eqnarray}
E\Bk{\eta(t)\eta^*(t')} = R\delta(t-t'),
\qquad
E\Bk{\eta(t)\eta(t')} = 0.
\label{eta}
\end{eqnarray}
We assume that the parameters 
\begin{eqnarray}
\theta = (\Omega, \gamma, A, C, R)^\top
\label{theta}
\end{eqnarray}
are constant in time, such that $z(t)$, $\xi(t)$, $y(t)$, and
$\eta(t)$ are stationary stochastic processes given $\theta$. $y(t)$,
in particular, has a power spectrum given by
\begin{eqnarray}
S_y(\omega|\theta) &\equiv&
\lim_{T\to\infty} E\Bk{\frac{1}{T} \abs{\int_{-T/2}^{T/2}dt y(t)\exp(-i\omega t)}^2} \label{Sy} \\
&=& AS(\omega) + R, \\
S(\omega) &\equiv& \frac{C^2}{(\omega-\Omega)^2+\gamma^2}.
\end{eqnarray}
Although this simple model suffices to describe our experiment, it is
not difficult to generalize our entire formalism to describe more
complicated dynamics and colored noise \cite{vantrees}. This is done
by generalizing $z(t)$ to a vector of state variables for more
mechanical and optical modes, Eq.~(\ref{a}) to a vectoral equation of
motion, and the parameters $(\Omega, \gamma, A, C, R)$ to matrices
that describe the coupled-mode dynamics and the noise statistics.

\subsection{Binary hypothesis testing}
Although hypothesis testing \cite{levy,vantrees3} is not the focus of
our study, the theory is useful for the derivation of the Cram\'er-Rao
bound, so we present the topic here briefly for completeness.

Suppose that there are two hypotheses, denoted by $\mathcal H_0$ and
$\mathcal H_1$, with prior probabilities $P_0$ and $P_1 = 1-P_0$. From
a measurement record $Y$, with a probability density $P(Y|\mathcal
H_0)$ or $P(Y|\mathcal H_1)$ that depends on the hypothesis, one
wishes to decide which hypothesis is true. Given the densities and a
decision rule, one can compute $P_{jk}$, the probability that
$\mathcal H_j$ is chosen when $\mathcal H_k$ is true. The average
error probability is
\begin{eqnarray}
P_e \equiv P_{10}P_0 + P_{01} P_1.
\end{eqnarray}
$P_e$ can be minimized using a Bayes likelihood-ratio test:
\begin{eqnarray}
\Lambda \equiv \frac{P(Y|\mathcal H_1)}{P(Y|\mathcal H_0)}
\overset{\mathcal H_1}{\underset{\mathcal H_0}{\gtrless}} \frac{P_0}{P_1},
\end{eqnarray}
which means that $\mathcal H_1$ is chosen if $\Lambda \ge P_0/P_1$ and
vice versa. The resulting $P_e$ is often difficult to compute
analytically, but can be bounded by upper and lower bounds. For $P_0 =
P_1 = 1/2$ \cite{levy,kailath},
\begin{eqnarray}
\frac{1}{2}\Bk{1-\sqrt{1-F^2(0.5)}}\le \min P_e 
\le
\frac{1}{2}\min_{0\le s\le 1} F(s),
\end{eqnarray}
where the upper bound is the Chernoff bound,
\begin{eqnarray}
F(s) \equiv E\Bk{\Lambda^s|\mathcal H_0} = 
\int dY P(Y|\mathcal H_0)
\Bk{\frac{P(Y|\mathcal H_1)}{P(Y|\mathcal H_0)}}^s,
\label{chernoff}
\end{eqnarray}
and $F(0.5)$ is known as the Bhattacharyya distance between the two
probability densities \cite{kailath}.

Let $Y$ be a record of continuous measurements:
\begin{eqnarray}
Y = \{x(t); -T/2\le t \le T/2\}.
\end{eqnarray}
If $x(t)$ is a realization of a real zero-mean stationary process with spectrum $S_x(\omega|\mathcal H_j)$, the exponent of $F(s)$ in the case of stationary processes and long observation time (SPLOT) is known to be \cite{levy}
\begin{eqnarray}
\Gamma_F &\equiv \lim_{T\to\infty} -\frac{1}{T}\ln F(s)
\\
&= \frac{1}{2}\intall \frac{d\omega}{2\pi}
\ln \frac{s S_x(\omega|\mathcal H_0) + (1-s) S_x(\omega|\mathcal H_1)}
{S_x^{s}(\omega|\mathcal H_0)S_x^{1-s}(\omega|\mathcal H_1)}.
\label{exponent}
\end{eqnarray}
This expression means that $F(s)$ has the form of
\begin{eqnarray}
F(s) = \beta(T)\exp\bk{-\Gamma_F T},
\end{eqnarray}
where $-\ln \beta(T)$ is asymptotically smaller than $T$:
\begin{eqnarray}
\lim_{T\to\infty} -\frac{1}{T}\ln \beta(T) = 0,
\end{eqnarray}
and therefore $\beta(T)$ decays more slowly than $\exp(-\Gamma_F T)$.

For the model in Sec.~\ref{model}, $y(t)$ is a complex signal, and (\ref{exponent}) needs to be modified. This can be done by assuming that $y(t)$ is bandlimited in $[-\pi b,\pi b]$ and considering a real signal $x(t)$ given by
\begin{eqnarray}
x(t) \equiv y(t)\exp(i\omega_0 t) + y^*(t)\exp(-i\omega_0t),
\end{eqnarray}
where $\omega_0$ is a carrier frequency assumed to be $> \pi b$. We then have
\begin{eqnarray}
S_x(\omega|\mathcal H_j) = S_y(\omega-\omega_0|\mathcal H_j) + 
S_y(-\omega-\omega_0|\mathcal H_j).
\end{eqnarray}
Using this expression in (\ref{exponent}) leads to another expression for the Chernoff exponent given by
\begin{eqnarray}
\Gamma_F
= \int_{-\pi b}^{\pi b} \frac{d\omega}{2\pi}
\ln \frac{s S_y(\omega|\mathcal H_0) + (1-s) S_y(\omega|\mathcal H_1)}
{S_y^s(\omega|\mathcal H_0)S_y^{1-s}(\omega|\mathcal H_1)}.
\label{exponent2}
\end{eqnarray}
Note the absence of the $1/2$ factor in (\ref{exponent2}) compared with (\ref{exponent}).

\subsection{Cram\'er-Rao Bound}
We now consider the estimation of $\theta$ from $Y$ with a probability
density given by $P(Y|\theta)$. Defining the estimate as $\hat
\theta(Y)$, the error covariance matrix is
\begin{eqnarray}
\Sigma(\theta) \equiv E\BK{\Bk{\hat \theta(Y)-\theta}
\Bk{\hat \theta(Y)-\theta}^\top\Big|\theta},
\\
E[g(Y)|\theta] \equiv \int dY P(Y|\theta) g(Y).
\end{eqnarray}
Assuming that $\hat \theta$ satisfies the unbiased condition
\begin{eqnarray}
E\Bk{\hat\theta(Y)\Big|\theta} = \theta,
\end{eqnarray}
the Cram\'er-Rao bound on $\Sigma$ is \cite{levy}
\begin{eqnarray}
\Sigma(\theta) \ge J^{-1}(\theta),
\end{eqnarray}
where $J(\theta)$ is known as the Fisher information matrix:
\begin{eqnarray}
J(\theta) \equiv E\BK{
\nabla\Bk{\ln P(Y|\theta)}
\nabla^\top\Bk{\ln P(Y|\theta)}\Big|\theta},
\\
\nabla \equiv \bk{\parti{}{\theta_1},\parti{}{\theta_2},\dots}^\top.
\end{eqnarray}
It turns out that $J(\theta)$ can be related to the Bhattacharyya
distance in a hypothesis testing problem with
\begin{eqnarray}
P(Y|\mathcal H_0) &= P(Y|\theta),
\\
P(Y|\mathcal H_1) &= P(Y|\theta').
\end{eqnarray}
$F(s)$ defined in (\ref{chernoff}) becomes a function of $\theta$ and
$\theta'$, and $J(\theta)$ can be expressed as \cite{vantrees3}
\begin{eqnarray}
  J(\theta) &= -4 \nabla \nabla^\top \ln
  F(0.5,\theta,\theta')\Big|_{\theta'=\theta}.
\label{J}
\end{eqnarray}
For the model in Section~\ref{model}, $y(t)$ is a realization of a
stationary process given $\theta$, so $J(\theta)$ for the estimation
of $\theta$ from $Y=\{y(t); -T/2 \le t \le T/2\}$ in the SPLOT case
can be obtained by combining (\ref{exponent2}) and (\ref{J}):
\begin{eqnarray}
\Gamma_J &\equiv \lim_{T\to\infty}\frac{J(\theta)}{T} 
\\
&= 4\nabla\nabla^\top\int_{-\pi b}^{\pi b} 
\frac{d\omega}{2\pi}
\ln \frac{S_y(\omega|\theta) + S_y(\omega|\theta')}
{2\sqrt{S_y(\omega|\theta)S_y(\omega|\theta')}}\Bigg|_{\theta'=\theta},
\end{eqnarray}
where $S_y(\omega|\theta)$ is given by (\ref{Sy}).  This expression
means that $J(\theta)$ for any stationary-process parameter estimation
problem increases linearly with time as $T\to\infty$, in the sense of
\begin{eqnarray}
J(\theta) &= \Gamma_J T +  o(T),
\end{eqnarray}
where $o(T)$ is asymptotically smaller than $T$:
\begin{eqnarray}
\lim_{T\to\infty} \frac{o(T)}{T} = 0.
\end{eqnarray}
In the asymptotic limit, maximum-likelihood (ML) estimation can attain
the Cram\'er-Rao bound \cite{shumway_stoffer}, so the bound is a
meaningful indicator of estimation error. Despite the asymptotic
assumption, the simpler analytic expressions are more convenient to
use for experimental design purposes.

Although the preceding formalism is applicable to the estimation of
any of the parameters, in the following we focus on $A$, the force
noise power. The Cram\'er-Rao bound on the mean-square estimation
error $\Sigma_A$ is
\begin{eqnarray}
\Sigma_A &\equiv E\BK{\Bk{\hat A(Y)-A}^2\Big|\theta}
\ge J_A^{-1},
\label{CRBA}
\\
\Gamma_A &\equiv \lim_{T\to\infty} \frac{J_A}{T} = 
\int_{-\pi b}^{\pi b}\frac{d\omega}{2\pi}
\frac{S^2(\omega)}{[AS(\omega)+R]^2}.
\label{GammaA}
\end{eqnarray}
This bound allows us to investigate the efficiency of the parameter
estimation algorithms presented in the next section.

\section{\label{algorithms}Parameter estimation algorithms}

\subsection{Averaging}


We first consider the estimator used in
Refs.~\cite{gavartin,harris_bowen2013}:
\begin{eqnarray}
\hat A_{\rm avg} = G\int_{-T/2}^{T/2} dt |y(t)|^2,
\qquad
G = \Bk{T\int_{-\pi b}^{\pi b} \frac{d\omega}{2\pi} S(\omega)}^{-1}.
\label{avg}
\end{eqnarray}
%
The rationale for this simple averaging estimator is that, in the absence of measurement noise ($R = 0$), it is an unbiased estimate for $T\to\infty$:
\begin{eqnarray}
\lim_{T\to\infty} E\bk{\hat A_{\rm avg}\Big|\theta,R = 0} = A.
\end{eqnarray}
The unbiased condition breaks down, however, in the presence of
measurement noise, and we are therefore motivated to find a better
estimator.

\subsection{Radiometer}

The ``radiometer'' estimator described in Ref.~\cite{vantrees3} can be
easily generalized for complex variables. The result is
\begin{eqnarray}
\hat A_{{\rm rad}} = G\Bk{\int_{-T/2}^{T/2}dt \int_{-T/2}^{T/2} dt' y^*(t) h(t-t') y(t') - B},
\end{eqnarray}
where $h(t-t')$ filters $y(t')$ before correlating the result with $y^*(t)$, and $G$ and $B$ are parameters chosen to enforce the unbiased condition.
We see that the averaging estimator $\hat A_{{\rm avg}}$ given by (\ref{avg}) also has the radiometer form. It can be shown that, for $T\to\infty$,
\begin{eqnarray}
G = \Bk{T\int_{-\pi b}^{\pi b} \frac{d\omega}{2\pi} H(\omega) S(\omega)}^{-1},
\\
B = T \int_{-\pi b}^{\pi b} \frac{d\omega}{2\pi} H(\omega) R,
\\
H(\omega) \equiv \intall dt h(t)\exp(-i\omega t).
\end{eqnarray}
The mean-square error, on the other hand, has the asymptotic expression
\begin{eqnarray}
\lim_{T\to\infty} \Sigma_A T = 
G^2\int_{-\pi b}^{\pi b} \frac{d\omega}{2\pi} H^2(\omega) S_y^2(\omega|\theta).
\end{eqnarray}
This expression coincides with the Cram\'er-Rao bound given by (\ref{CRBA}) and (\ref{GammaA}) if we set
\begin{eqnarray}
H(\omega) = \frac{S(\omega)}{[DS(\omega)+R]^2},
\label{H}
\end{eqnarray}
and $A$ happens to be equal to $D$. For any other value of $A$, the radiometer is suboptimal.

\subsection{Expectation-maximization (EM) algorithm}
A major shortcoming of the radiometer is its requirement of parameters
other than $A$ to be known exactly. Another issue is that it assumes
continuous time and relies on asymptotic arguments, when the
measurements are always discrete and finite in practice.  We find that
the EM algorithm \cite{dempster,shumway_stoffer,levy}, which performs
maximum-likelihood (ML) estimation and is applicable to the linear
Gaussian model we consider here, overcomes both of these problems.


ML estimation aims to find the set of parameters $\theta$ that
maximizes the log-likelihood function $\ln P(Y|\theta)$. This task can
be significantly simplified by the EM algorithm if there exist hidden
data $Z$ that results in simplified expressions for $P(Z|Y,\theta)$
and $P(Y,Z|\theta)$.  Starting with a trial $\theta = \theta^{0}$, the
algorithm considers the estimated log-likelihood function
\begin{eqnarray}
Q(\theta,\theta^{k}) \equiv \int dZ P(Z|Y,\theta^{k})
\ln P(Y,Z|\theta),
\end{eqnarray}
where the superscript $k$ is an index denoting the EM iteration, and
finds the $\theta^{k+1}$ for the next iteration by maximizing $Q$:
\begin{eqnarray}
\theta^{k+1} = \arg \max_{\theta} Q(\theta,\theta^{k}).
\end{eqnarray}
The iteration is halted when the difference betwen $\theta^{k+1}$ and
$\theta^{k}$ reaches a prescribed threshold, and the final
$\theta^{k+1}$ is taken to be the EM estimate
$\hat\theta_{{\rm EM}}$.

To apply the EM algorithm to our model in Section~\ref{model}, we
consider a complex discrete-time Gauss-Markov model:
\begin{eqnarray}
z_{j+1} = {f} z_j + w_j,
\label{z}\\
y_j = c z_j + v_j,
\quad
j = 0,1,\dots,J.
\label{Y}
\end{eqnarray}
In general, $z_j$ and $y_j$ can be column vectors, and ${f}$ and $c$
are matrices. $w_j$ and $v_j$ are complex independent zero-mean
Gaussian random variables with covariances given by
\begin{eqnarray}
E\bk{w_jw_k^\dagger} = q\delta_{jk},
\qquad
E\bk{w_jw_k^\top} = 0,
\label{w}
\\
E\bk{v_jv_k^\dagger} = r\delta_{jk},
\qquad
E\bk{v_jv_k^\top} = 0,
\label{v}
\end{eqnarray}
where $^\dagger$ denotes the conjugate transpose, $^\top$ denotes the
transpose, and $q$ and $r$ are covariance matrices. The parameters of
interest $\theta$ are the components of ${f}$, $c$, $q$, and $r$. The
EM algorithm for a real Gauss-Markov model described in
Refs.~\cite{shumway_stoffer,levy} is generalized to account for
complex variables in \ref{complexEM}. The problem may become
ill-conditioned when too many parameters are taken to be unknown and
multiple ML solutions exist \cite{shumway_stoffer,levy,guta_yamamoto},
so we choose a parameterization with known $q$:
\begin{eqnarray}
{f} &= \exp\Bk{\bk{i\Omega-\gamma} \delta t},
\\
c &= C\sqrt{A\frac{1-\exp(-2\gamma\delta t)}{2\gamma \delta t}},
\\
q &= \delta t,
\\
r &= \frac{R}{\delta t},
\end{eqnarray}
where $\delta t$ is the sampling period. With the EM estimates $\hat
f_{{\rm EM}}$, $\hat c_{{\rm EM}}$, and $\hat r_{{\rm EM}}$
and assuming that $\delta t$ and $C$ are known by independent
calibrations, we can retrieve estimates of $\Omega$, $\gamma$, $A$,
and $R$:
\begin{eqnarray}
\hat \Omega_{{\rm EM}} &= \frac{\arg\hat f_{{\rm EM}}}{\delta t},
\\
\hat\gamma_{{\rm EM}} &= -\frac{\ln |\hat f_{{\rm EM}}|}{\delta t},
\\
\hat A_{{\rm EM}} &= \frac{\hat c_{{\rm EM}}^2}{C^2}
\frac{2\hat\gamma_{{\rm EM}}\delta t}
{1-\exp(-2\hat\gamma_{{\rm EM}}\delta t)},
\\
\hat R_{{\rm EM}} &= \hat r_{{\rm EM}} \delta t.
\end{eqnarray}
It can be shown that the ML parameter estimator for the Gauss-Markov
model is asymptotically efficient \cite{shumway_stoffer}, meaning that
it attains the Cram\'er-Rao bound in the limit of $T\to\infty$.

\section{Application to experimental data}

\subsection{Procedure}
There are two records of experimental data, one with thermal noise in
$\xi(t)$ and one with additional applied white noise in $\xi(t)$,
leading to a different $A$ for each record, denoted by $A^{(0)}$ and
$A^{(1)}$. Each record contains $J_{{\rm max}}+1=3,750,001$ points
of $y_j^{(n)}$. With a sampling frequency $b = 1/\delta t = 15~$MHz,
the total time for each record is $T_{{\rm max}} =
(J_{{\rm max}} + 1)\delta t \approx 0.25$~s. From independent
calibrations, we also obtain $C = 2.61\times
10^{-2}~({\rm fN}/\sqrt{{\rm Hz}})^{-1}$. To investigate the
errors with varying $T$, we divide each record into slices of records
with various $T$, resulting in $M(T) =
{\rm floor}(T_{{\rm max}}/T)$ number of trials for each
$T$.  Using a desktop computer (Intel Core i7-2600 CPU@3.4GHz with
16GB RAM) and MATLAB, we apply each of the three estimators in
Section~\ref{algorithms} to each trial to produce an estimate $\hat
A_{m,l}^{(n)}(T)$, where $m$ denotes the trial and $l$ denotes the
estimator.  The EM iteration is stopped when the fractional difference
between the current estimate of $A$ and the previous value is less
than $10^{-7}$. For the averaging and radiometer estimators, true
values for $\Omega$, $\gamma$, and $R$ are needed, and since we do not
know them, we estimate them by applying the EM algorithm to the whole
records. This is reasonable because $T_{{\rm max}} \gg
4~{\rm ms} \ge T$, and we expect
$\hat\theta_{{\rm EM}}^{(n)}(T_{{\rm max}})$ to be much closer
to the true values $\theta^{n}$ than the short-time estimates. The EM
algorithm for each $T$, on the other hand, does not use $\hat
\theta_{{\rm EM}}^{(n)}(T_{{\rm max}})$ at all and produces its
own estimates each time. The parameter $D$ in (\ref{H}) is taken to be
$\hat A_{{\rm EM}}^{(0)}(T_{{\rm max}})$.  The estimation errors
are computed by
\begin{eqnarray}
\Sigma_l^{(n)}(T) &= \frac{1}{M(T)}\sum_{m=1}^{M(T)} \Bk{\hat A_{m,l}^{(n)}(T) -A^{(n)}}^2,
\end{eqnarray}
and compared with the SPLOT Cram\'er-Rao bound $J_A^{-1} \approx
(\Gamma_J T)^{-1}$ by assuming $\theta^{(n)} = \hat\theta_{
  {\rm EM}}^{(n)} (T_{{\rm max}})$.

Note that the estimation error in general contains two components:
\begin{eqnarray}
\Sigma &= \frac{1}{M}\sum_{m=1}^{M} \bk{\hat A_{m} -\bar A}^2 +
\bk{\bar A-A}^2,
\end{eqnarray}
where
\begin{eqnarray}
\bar A &\equiv \frac{1}{M}\sum_{m=1}^{M} \hat A_{m}
\end{eqnarray}
is the sample mean of the estimate, the first component is the sample
variance, and the second component is the square of the estimate
\emph{bias} with respect to the true value $A$. Unlike
Refs.~\cite{gavartin,harris_bowen2013}, our error analysis is able to
account for the bias component more accurately by referencing with the
much more accurate long-time EM estimates.


\subsection{Results}
Applied to the two records, the EM algorithm produces the following
estimates: 
\begin{eqnarray}
\hat A_{{\rm EM}}^{(0)}(T_{\rm max}) &=& 2.4748/C^2 
=3.64\times 10^{3}~{\rm fN}^2\,{\rm Hz}^{-1},
\\
\hat \Omega_{{\rm EM}}^{(0)}(T_{{\rm max}}) &=& -1.8582\times 10^{4}~{\rm rad~s}^{-1},
\\
\hat \gamma_{{\rm EM}}^{(0)}(T_{{\rm max}}) &=& 5.5730\times 10^{4}~{\rm rad~s}^{-1},
\\
\hat R_{{\rm EM}}^{(0)}(T_{{\rm max}}) &=& 1.4532\times 10^{-13}~{\rm Hz}^{-1},
\\
\hat A_{{\rm EM}}^{(1)}(T_{\rm max}) &=& 2.6926/C^2
= 3.96\times 10^{3}~{\rm fN}^2\,{\rm Hz}^{-1},
\\
\hat \Omega_{{\rm EM}}^{(1)}(T_{{\rm max}}) &=& -1.8668\times 10^{4}~{\rm rad~s}^{-1},
\\
\hat \gamma_{{\rm EM}}^{(1)}(T_{{\rm max}}) &=& 5.6156\times 10^{4}~{\rm rad~s}^{-1},
\\
\hat R_{{\rm EM}}^{(1)}(T_{{\rm max}}) &=& 1.4703\times 10^{-13}~{\rm Hz}^{-1}.
\end{eqnarray}
The algorithm takes $\approx 3.3$ hours to run for each record.  These
values are then used as references to analyze the estimators at
shorter times.

\begin{figure}[htbp]
\centerline{\includegraphics[width=\textwidth]{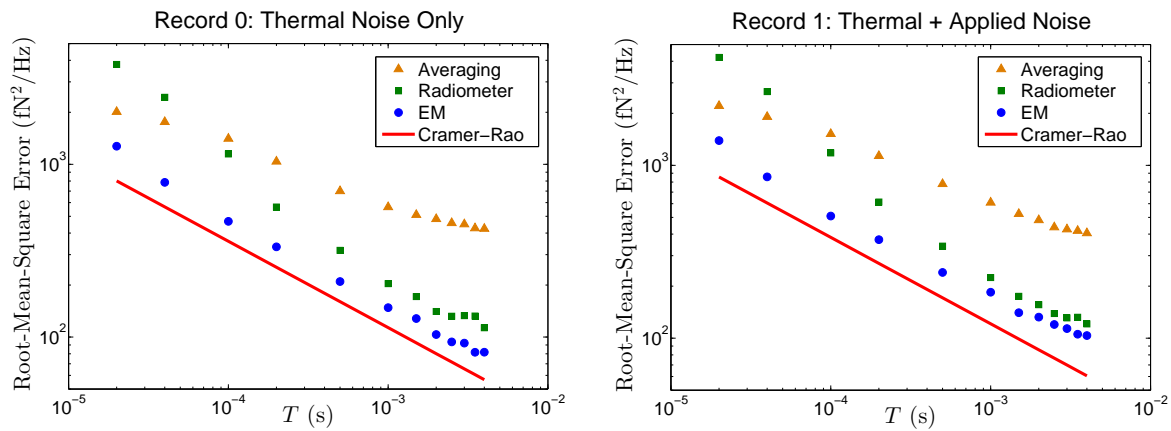}}
\caption{
  Root-mean-square
  force-noise-power estimation errors and the asymptotic Cram\'er-Rao
  bound versus time in log-log scale. Left: the force contains
  thermal noise only. Right: the force contains thermal noise and an
  applied noise.}
\label{MSE_combined}
\end{figure}

Figure~\ref{MSE_combined} plots the root-mean-square errors
$\sqrt{\Sigma_l^{(n)}(T)}$ and the SPLOT Cram\'er-Rao bound
$J_A^{-1/2} \to (\Gamma_JT)^{-1/2}$ versus time $T$ in log-log
scale. The two plots show very similar behavior. A few observations
can be made:
\begin{enumerate}
\item The averaging estimator is more accurate than the radiometer for
  short times but becomes much worse for longer times. We cannot
  explain the short-time errors because our analytic results rely on
  the long-time limit, although the errors there are so high relative
  to the estimate that they are irrelevant to real applications. The
  large long-time errors can be attributed to the bias and
  suboptimality of the estimator.

\item The radiometer beats the averaging estimator and approaches the
  Cram\'er-Rao bound for longer times. This is consistent with our
  SPLOT analysis, as we have chosen $D = \hat
  A_{{\rm EM}}^{(0)}(T_{{\rm max}})$ and the radiometer should
  be near-optimal.

\item The EM estimator beats the other estimators at all times and
  follow the Cram\'er-Rao bound more closely, even though we allow the
  averaging and radiometer estimators to have the unfair advantage of
  accessing more accurate values of $\Omega$, $\gamma$, and $R$. This
  may be explained by the fact that the EM algorithm is formulated to
  perform ML estimation on discrete measurements for any finite $T$,
  unlike the other estimators that rely only on asymptotic arguments.
  
\item The EM estimator takes a much longer time to compute
  (computation time $\approx 200~$s for one trial with $J+1 = 60,000$
  points and $T = 4~$ms) than the other estimators ($\approx 0.3$~ms
  for the averaging estimator, $\approx 16$~ms for the radiometer). If
  computation time is a concern, the radiometer estimator may be
  preferable, although its performance depends heavily on the accuracy
  of the other assumed parameters, and the EM method can still be
  useful for estimating such parameters in offline system
  identification.

\end{enumerate}

\begin{figure}[htbp]
\centerline{\includegraphics[width=\textwidth]{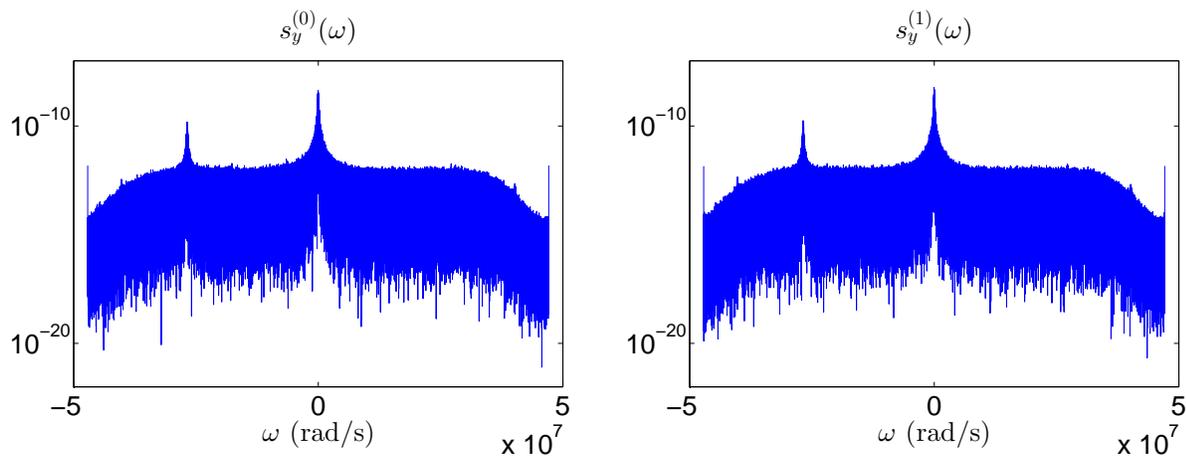}}
  \caption{Raw spectra $s_y^{(n)}(\omega)$ of
    the measurement records $y_j^{(n)}$ in log scale.}
\label{raw_spectra}
\end{figure}

To gain further insight into the finite gap between the errors and the
Cram\'er-Rao bound, in Figure~\ref{raw_spectra} we plot the raw
spectrum of $y_j^{(n)}$, defined as
\begin{eqnarray}
s_y^{(n)}(\omega) \equiv \frac{1}{T_{{\rm max}}}
\abs{\delta t\sum_{j=0}^{J_{{\rm max}}} y_j^{(n)} 
\exp(-i\omega j\delta t)}^2.
\end{eqnarray}

The figure shows that our model does not exactly match the experiment
in two ways:
\begin{enumerate}
\item The data show a second weaker resonance peak.
\item The noise floor of the data rolls off at higher frequencies
due to the presence of an RF notch filter in the experiment
prior to data acquisition.
\end{enumerate}
Despite the mismatch, our results are in reasonable agreement with the
theory.  To improve the estimation accuracy further, the weaker
resonance can be modeled by including another mode in our linear
Gaussian model, while the noise-floor roll-off can be removed by a
whitening filter before applying the estimators.

\section{Outlook}
In this paper we have followed the paradigm of orthodox statistics to
investigate parameter estimation for an optomechanical system,
focusing on unbiased and ML estimators and the Cram\'er-Rao bound. For
detection applications \cite{ting_hero_rugar} with uncertain
parameters, the ML estimator can form the basis of more advanced
hypothesis testing techniques, such as the generalized
likelihood-ratio test \cite{levy}. The assumption of static parameters
means that the presented techniques are most suited to system
identification purposes. For sensing applications, the parameters are
often time-varying, and Bayesian estimators, such as the extended and
unscented Kalman filters for continuous variables \cite{simon}, the
generalized-pseudo-Bayesian and interacting-multiple-model algorithms
for finite-state dynamical hypotheses \cite{bar-shalom}, and particle
filtering \cite{particle}, may be more suitable.

Since the Gauss-Markov model often remains valid for quantum systems
\cite{wiseman_milburn}, a quantum extension of our study is
straightforward. This means that the presented techniques are
potentially useful for future quantum sensing and system
identification applications, such as optomechanical force sensing
\cite{chen2013,wheatley,yonezawa,iwasawa}, atomic magnetometry
\cite{stockton,petersen}, and fundamental tests of quantum mechanics
\cite{chen2013,aspelmeyer2013,testing_quantum}. We expect our
parametric methods to lead to more accurate quantum sensing and
control than robust quantum control methods \cite{stockton,james},
which may be too conservative for the highly controlled environment of
typical quantum experiments.
There also exist quantum versions of the Cram\'er-Rao bound that
impose fundamental limits to the parameter estimation accuracy for a
quantum system with any measurement \cite{helstrom,twc,tsang_open},
and it may be interesting to explore how close the classical bounds
presented here can get to the quantum limits.
%

The continued improvement of optomechanical devices for applications
and fundamental science requires precise engineering of the mechanical
resonance frequency, dissipation rate and effective mass. This
necessitates a deep understanding of how these mechanical properties
depend on differing materials and fabrication techniques. The
mechanical resonance frequency is easily predicted via a numerical
eigenmode analysis using the geometry of the structure and the Youngs
modulus of the material. It is much more challenging to predict the
level of mechanical dissipation, where numerical models are not as
well established and multiple decay channels usually exist. Effective
experimental characterization of such dissipation channels requires
high precision force estimation to accurately quantify the oscillators
coupling to the environment. This is critical to advancing
optomechanics in applications such as quantum memories and quantum
information~\cite{Bagheri11_NatNano, Rabl10_NatPhys}.  A more
immediate application for high precision force estimation is that of
temperature sensing and bolometry where small relative changes of the
signal power are of interest, for example, in detecting submillimeter
wavelengths in radio astronomy~\cite{Griffin00_NucInst} or even to
search for low energy events in particle
physics~\cite{Alessandrello98_PRB}.  Given the demonstrated success of
our statistical techniques, we envision them to be similarly useful
for all these applications.


\section*{Acknowledgments}
  S.~Z.~Ang and M.~Tsang acknowledge support by the Singapore National
  Research Foundation under NRF Grant.~No.~NRF-NRFF2011-07.
  W.~P.~Bowen and G.~I.~Harris acknowledge funding from the Australian
  Research Council Centre of Excellence CE110001013 and Discovery
  Project DP0987146. Device fabrication was undertaken within the
  Queensland Node of the Australian Nanofabrication Facility.

\appendix
\section{\label{complexEM}EM algorithm for the complex Gauss-Markov model}


The model of interest is described by (\ref{z})--(\ref{v}). The
parameters of interest, denoted by $\theta$, are the components of
$f$, $c$, $q$, and $r$. Generalizing the algorithm described in
\cite{shumway_stoffer,levy} for complex variables, we have
\begin{eqnarray}
-\ln P(Y,Z|\theta) 
= \sum_{j=0}^{J-1}\bk{z_{j+1}-{f}z_j}^\dagger q^{-1}\bk{z_{j+1}-{f}z_j}
+J\ln \det q  
\nonumber\\
+ \sum_{j=0}^J \bk{y_j - cz_j}^\dagger r^{-1} \bk{y_j - cz_j}+ 
(J+1)\ln \det r + \alpha,
\end{eqnarray}
where $\alpha$ does not depend on $\theta$ and is discarded. To
compute the estimated log-likelihood function $Q(\theta,\theta^{k})$,
we need
\begin{eqnarray}
\hat z_j^{k} &\equiv& E(z_j|Y,\theta^{k}),
\label{zhat}\\
\epsilon_j^{k} &\equiv& z_j - \hat z_j^{k},
\label{epsilon}\\
\Pi_j^{k} &\equiv& E(\epsilon_j^{k}\epsilon_j^{k\dagger} |Y,\theta^{k}),
\label{Pi}\\
\Pi_{j,j-1}^{k} &\equiv& E(\epsilon_j^{k} \epsilon_{j-1}^{k\dagger} |Y,\theta^{k}),
\label{Pi2}
\end{eqnarray}
which can be computed by the Rauch-Tung-Striebel (RTS) smoother
\cite{levy,simon}.  Starting with stationary initial conditions for
$\hat z_{-1}^{+k}$ and $\Pi_{-1}^{+k}$, the smoother consists of a
forward Kalman filter:
\begin{eqnarray}
\hat z_{j}^{-k} &=& {f}^{k} \hat z_{j-1}^{+k},
\\
\Pi_{j}^{-k} &=& {f}^{k} \Pi_{j-1}^{+k}{f}^{k\dagger} + q,
\\
K_j^{+k} &=& \Pi_j^{-k}c^{k\dagger}\bk{c^{k}\Pi_j^{-k}c^{k\dagger} + r^{k}}^{-1},
\\
\hat z_j^{+k} &=& \hat z_j^{-k} + K_j^{+k}\bk{y_j- c^{k} \hat z_j^{-k}},
\\
\Pi_j^{+k} &=& \bk{I-K_j^{+k}c^{k} } \Pi_j^{-k}\bk{I-K_j^{+k}c^{k}}^\dagger 
+ K_j^{+k} r^{k}K_j^{+k\dagger},
\end{eqnarray}
until $j = J$, and a backward propagation:
\begin{eqnarray}
\hat z_J^{k} &=& \hat z_J^{+k},
\\
\Pi_J^{k} &=& \Pi_J^{+k},
\\
K_j^{k} &=& \Pi_j^{+k} {f}^{k\dagger}\bk{\Pi_{j+1}^{-k}}^{-1},
\\
\hat z_j^{k} &=& \hat z_j^{+k} + K_j^{k}\bk{\hat z_{j+1}^{k}-\hat z_{j+1}^{-k}},
\\
\Pi_j^{k} &=& \Pi_j^{+k} - K_j^{k}\bk{\Pi_{j+1}^{-k}-\Pi_{j+1}^{k}}K_j^{k\dagger},
\\
\Pi_{j,j-1}^{k} &=& \Pi_j^{k} K_{j-1}^{k\dagger},
\end{eqnarray}
until $j = 0$. We can then write $Q(\theta,\theta^{k})$ as
\begin{eqnarray}
-Q(\theta,\theta^{k}) 
&= \trace\Bigg\{ q^{-1}\bk{\Phi^{k}-{f}\Psi^{k\dagger}-\Psi^{k} {f}^\dagger + 
{f}\Theta^{k} {f}^\dagger} + J\ln q 
\nonumber\\&\quad
+
r^{-1}\bk{\Upsilon-c\Xi^{k\dagger}-\Xi^k c^\dagger+c\Delta^k c^\dagger}
+ (J+1)\ln r\Bigg\},
\end{eqnarray}
where we have defined
\begin{eqnarray}
\Phi^{k} &\equiv& \sum_{j=1}^J \bk{\hat z_j^{k}\hat z_j^{k\dagger} + \Pi_j^{k}},
\\
\Psi^{k} &\equiv& \sum_{j=1}^J \bk{\hat z_j^{k} \hat z_{j-1}^{k\dagger} + \Pi_{j,j-1}^{k}},
\\
\Theta^{k} &\equiv& \sum_{j=0}^{J-1} \bk{\hat z_j^{k}\hat z_j^{k\dagger} + \Pi_j^{k}},
\\
\Upsilon &\equiv& \sum_{j=0}^J y_j y_j^\dagger,
\\
\Xi^k &\equiv& \sum_{j=0}^J y_j \hat z_j^{k\dagger},
\\
\Delta^k &\equiv& \sum_{j=0}^J \bk{\hat z_j^{k}\hat z_j^{k\dagger} + \Pi_j^{k}}.
\end{eqnarray}
Maximizing $Q(\theta,\theta^{k})$ with respect to $\theta$, we find 
\begin{eqnarray}
{f}^{k+1} &=& \Psi^{k}\bk{\Theta^{k}}^{-1},
\\
c^{k+1} &=& \Xi^k\bk{\Delta^k}^{-1},
\label{complex_c}
\\
q^{k+1} &=& \frac{1}{J}\Bk{\Phi^k-\Psi^{k}\bk{\Theta^{k}}^{-1}\Psi^{k\dagger}},
\label{qk1}\\
r^{k+1} &=& 
\frac{1}{J+1}
\Bk{\Upsilon-\Xi^k\bk{\Delta^k}^{-1}\Xi^{k\dagger}}.
\end{eqnarray}
One can simply take the real part of Eq.~(\ref{complex_c}) if $c$ is
known to be real. The complex EM algorithm turns out to be the same as
the real version with all transpose operations $^\top$ replaced by
conjugate transpose $^\dagger$.

The same algorithm is also applicable to the quantum Gauss-Markov
model \cite{wiseman_milburn}, as the RTS smoother is equivalent to the
linear quantum smoother \cite{petersen,smooth,smooth_pra1}. The
possibility of using the EM algorithm for quantum systems is also
mentioned in Ref.~\cite{gammelmark2013}. The complex model is more
compact when the noises are phase-insensitive. With phase-sensitive
noises, there is no computational advantage with a complex model and
one can just use a real model to describe the real and imaginary parts
separately.

\section*{References}
\bibliographystyle{hunsrt}
\bibliography{optomech_parameter_estimation3}
\end{document}